\documentclass[runningheads]{llncs}

\usepackage[english]{babel} 
\usepackage[utf8]{inputenc} 
\usepackage[T1]{fontenc}    

\usepackage{graphicx}       

\usepackage{amsmath}
\usepackage{amsfonts}
\usepackage{graphicx}
\usepackage{todonotes}
\usepackage{float}
\usepackage{listofitems}
\usepackage{tikz}
\usepackage{subcaption}
\usetikzlibrary{shapes,arrows,calc, fit, backgrounds}
\usepackage{comment}        

\usepackage{tabularx}
\usepackage{booktabs}

\usepackage[open,numbered]{bookmark} 

\usepackage[nocompress]{cite} 

\usepackage{hyperref}       

\usepackage[noabbrev,nameinlink]{cleveref}
\usepackage[acronym]{glossaries}

\begin{document}

\newacronym{al}{AL}{active learning}
\newacronym{dl}{DL}{deep learning}
\newacronym{fsl}{FSL}{few shot learning}
\newacronym{hci}{HCI}{human-computer interaction}
\newacronym{iml}{IML}{interactive machine learning}
\newacronym{iui}{IUI}{intelligent user interface}
\newacronym{ml}{ML}{machine learning}
\newacronym{pam}{PAM}{passive acoustic monitoring}
\newacronym{pa}{PA}{protected area}
\newacronym{pca}{PCA}{principal components analysis}
\newacronym{ssl}{SSL}{self-supervised learning}
\newacronym{vae}{VAE}{variational autoencoder}
\newacronym{vr}{VR}{virtual reality}

\title{A Virtual Reality Tool for Representing, Visualizing and Updating Deep Learning Models}

\titlerunning{Representing, Visualizing and Updating Deep Learning Models}
%

\newcommand{\repeatthanks}{\textsuperscript{\thefootnote}}

\author{Hannes Kath\thanks{Both authors contributed equally to this research.}\inst{1,2} \and
Bengt Lüers\repeatthanks\inst{1} \and
Thiago S. Gouvêa\inst{1,2} \and
Daniel Sonntag\inst{1,2}}

\authorrunning{Kath and Lüers, et al.}

\institute{
German Research Center for Artificial Intelligence (DFKI), Oldenburg, Germany \email{\{hannes\_berthold.kath, bengt.lueers, thiago.gouvea, daniel.sonntag\}@dfki.de}
\and
University of Oldenburg, Applied Artificial Intelligence (AAI), Oldenburg, Germany
}

%
\maketitle              

\begin{abstract}

Deep learning is ubiquitous, but its lack of transparency limits its impact on several potential application areas.
We demonstrate a virtual reality tool for automating the process of assigning data inputs to different categories.
A dataset is represented as a cloud of points in virtual space.
The user explores the cloud through movement and uses hand gestures to categorise portions of the cloud.
This triggers gradual movements in the cloud: points of the same category are attracted to each other, different
groups are pushed apart, while points are globally distributed in a way that utilises the entire space.
The space, time, and forces observed in virtual reality can be mapped to well-defined machine learning concepts, namely the latent space, the training epochs and the backpropagation.
Our tool illustrates how the inner workings of deep neural networks can be made tangible and transparent.
We expect this approach to accelerate the autonomous development of deep learning applications by end users in novel areas.


\end{abstract}

\glsresetall

\section{Introduction}

\Gls{ml} with deep neural networks, or \gls{dl}, has achieved astonishing performance in many tasks \cite{lecun2015deep}, and systems based on \gls{dl} are ubiquitous in our everyday lives.
However, for most people these systems are black boxes---the algorithms powering them are not transparent, understandable, or even approachable.
This lack of transparency raises ethical concerns \cite{EU2019AI} and limits the potential impact of \gls{ml} on several novel applications.
\Gls{iml} is the design and implementation of algorithms and \gls{iui} frameworks that facilitate \gls{ml} with the help of human interaction, and
includes the mission to
empower end users to develop their own domain-specific \gls{dl} applications \cite{Zacharias2018IML, simard2017machine}.

We demonstrate a \gls{vr} tool for automating the common supervised \gls{ml} task of assigning category labels to data inputs (e.g. classifying images).
Traditionally,
such \gls{ml} tasks would be implemented through a pipeline that starts with data annotation, followed by model design, training, and finally deployment.
Data annotation is the human-labor-intensive task of adding metadata (e.g. category labels) to a dataset with the purpose of providing examples to guide the training of an expert-designed \gls{ml} model.
The training process should render the model capable of generating sufficiently accurate category labels for previously unseen data inputs.
At this stage,
the resulting model can be deployed to power a user-facing system.
In such a traditional system, end user interaction with the system is limited to providing input data and collecting back a prediction.
While such black-box interaction patterns might suffice for many purposes, we propose an alternative interaction paradigm.

\vspace{-0.1cm}
\begin{figure}
    \centering
    \begin{subfigure}{0.49\textwidth}
        \centering
        \includegraphics[width=\textwidth, clip]{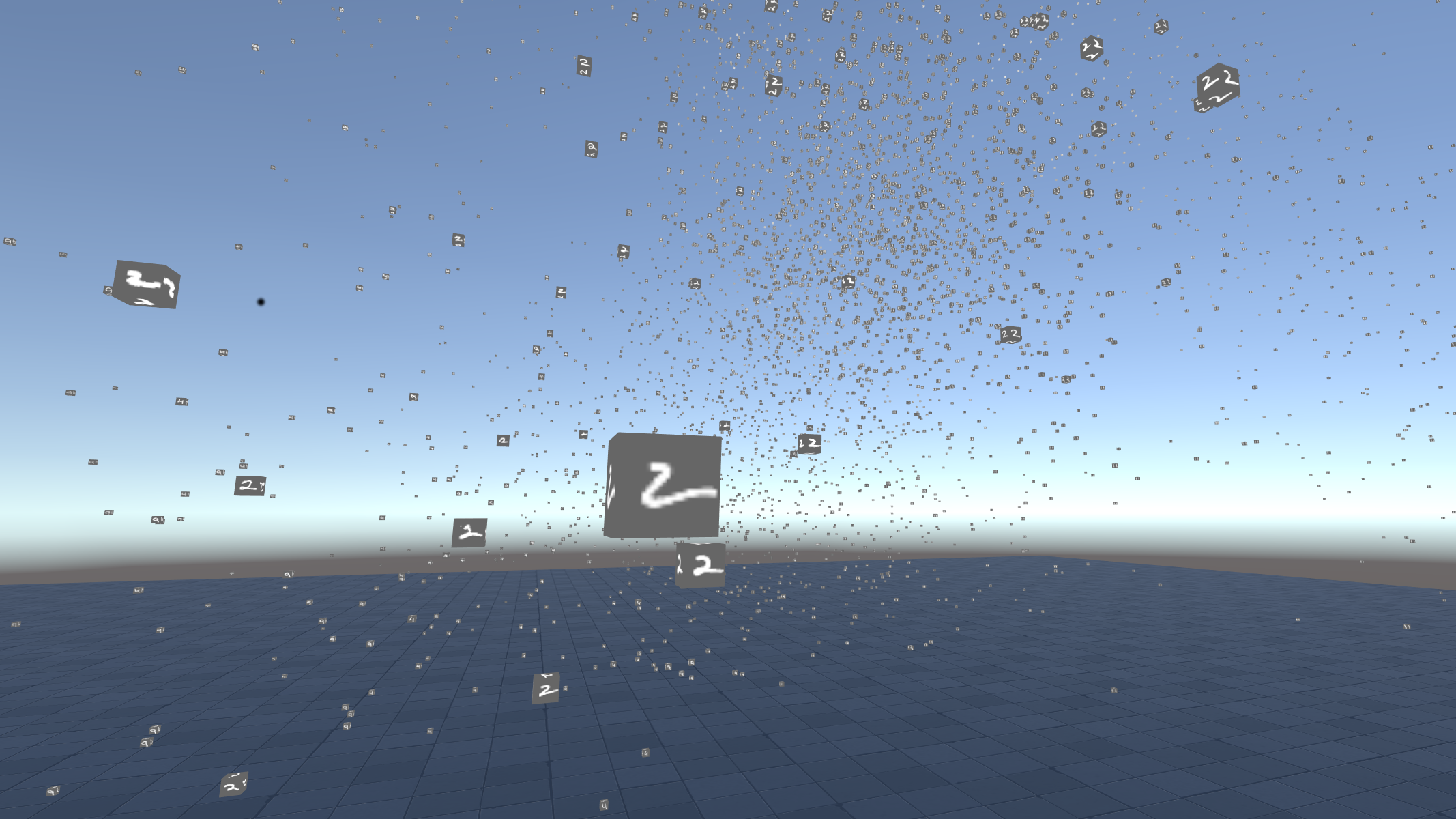}
        \caption{Initial unlabeled dataset, viewed from outside the cloud}
        \label{fig:InitDataComplete}
    \end{subfigure}
    \hspace{.0\textwidth}
    \begin{subfigure}{0.49\textwidth}
       \centering
       \includegraphics[width=\textwidth, clip]{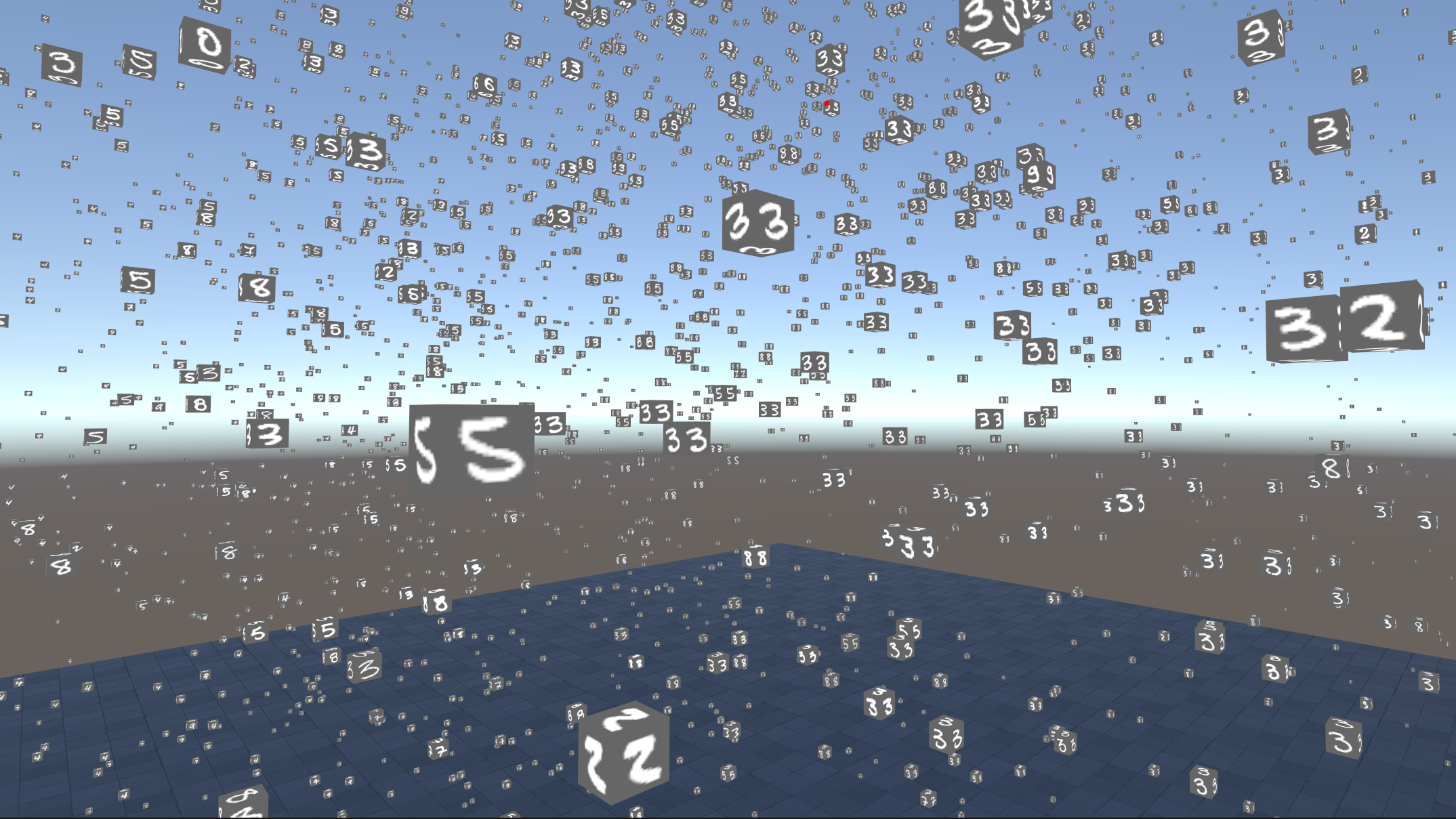}
       \caption{Initial unlabeled dataset, viewed from inside the cloud}
       \label{fig:InitDataZoomed}
    \end{subfigure}
    \hspace{.0\textwidth}
    \begin{subfigure}{0.49\textwidth}
        \centering
        \includegraphics[width=\textwidth, clip]{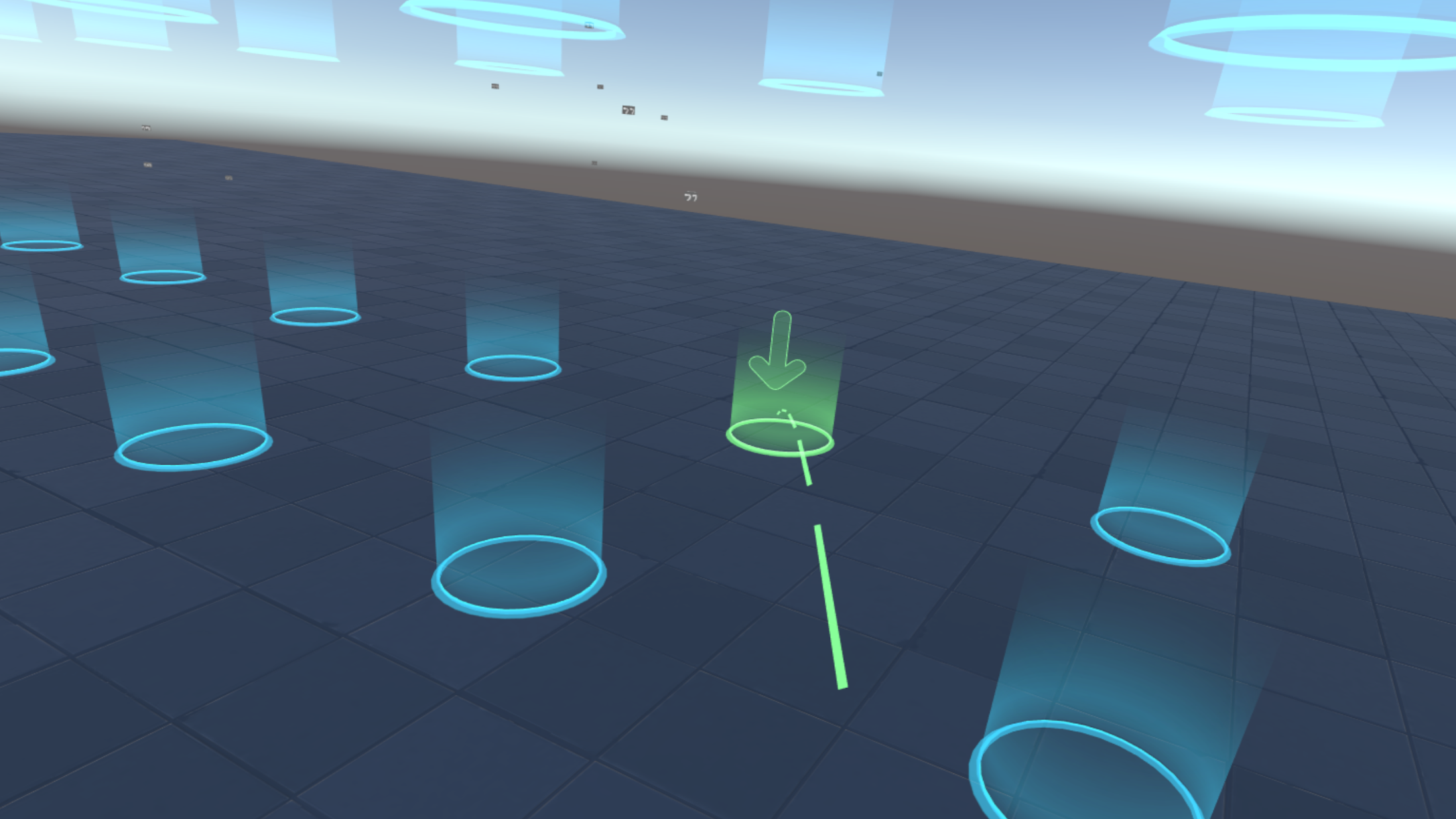}
        \caption{Teleport options for movement over long distances}
        \label{fig:Teleport}
   \end{subfigure}
   \hspace{.0\textwidth}
   \begin{subfigure}{0.49\textwidth}
        \centering
        \includegraphics[width=\textwidth, clip]{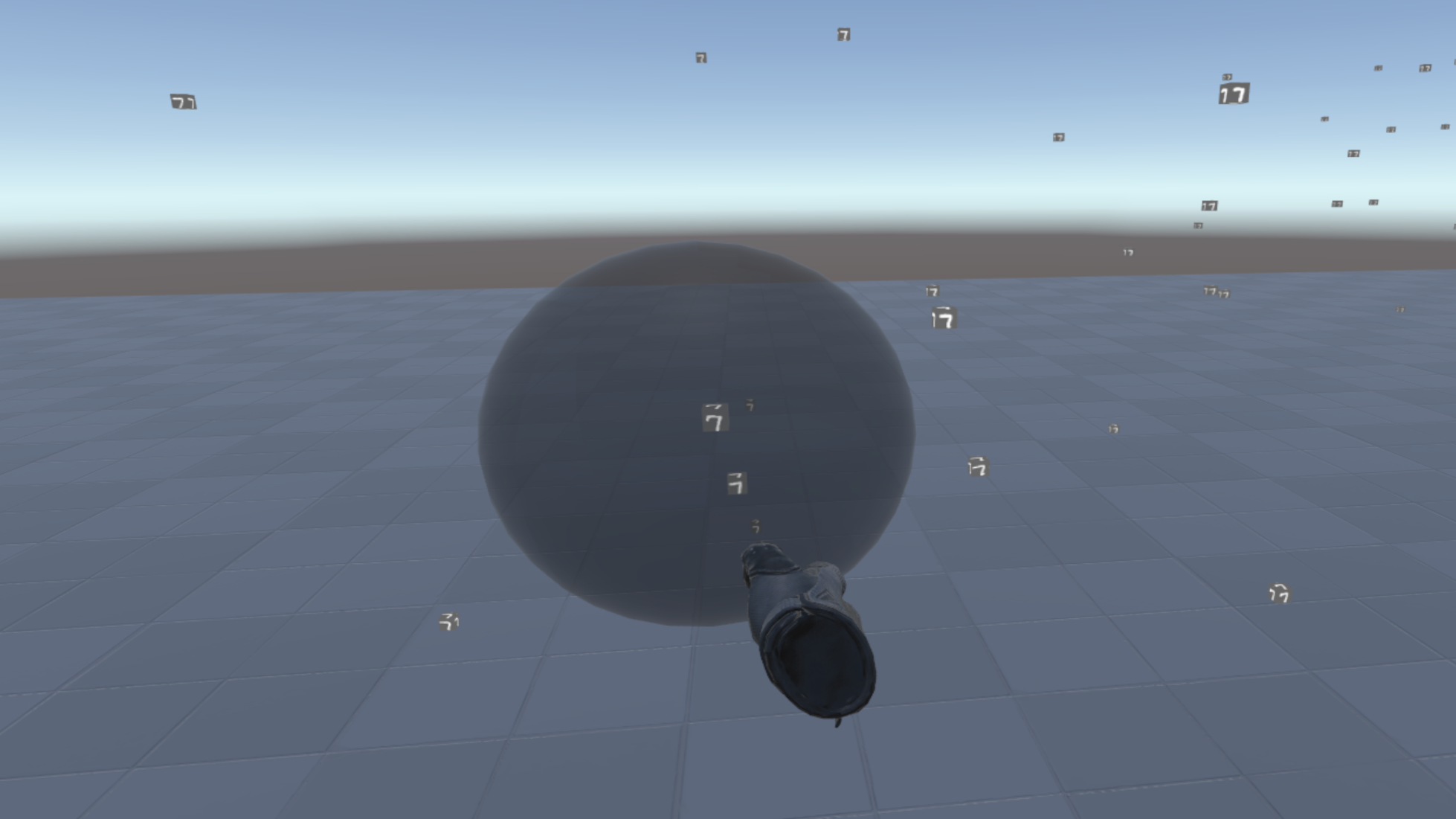}
        \caption{Positioning and sizing of a sphere for annotation}
        \label{fig:SpherePosition}
   \end{subfigure}
   \hspace{.0\textwidth}
   \begin{subfigure}{0.49\textwidth}
        \centering
        \includegraphics[width=\textwidth, clip]{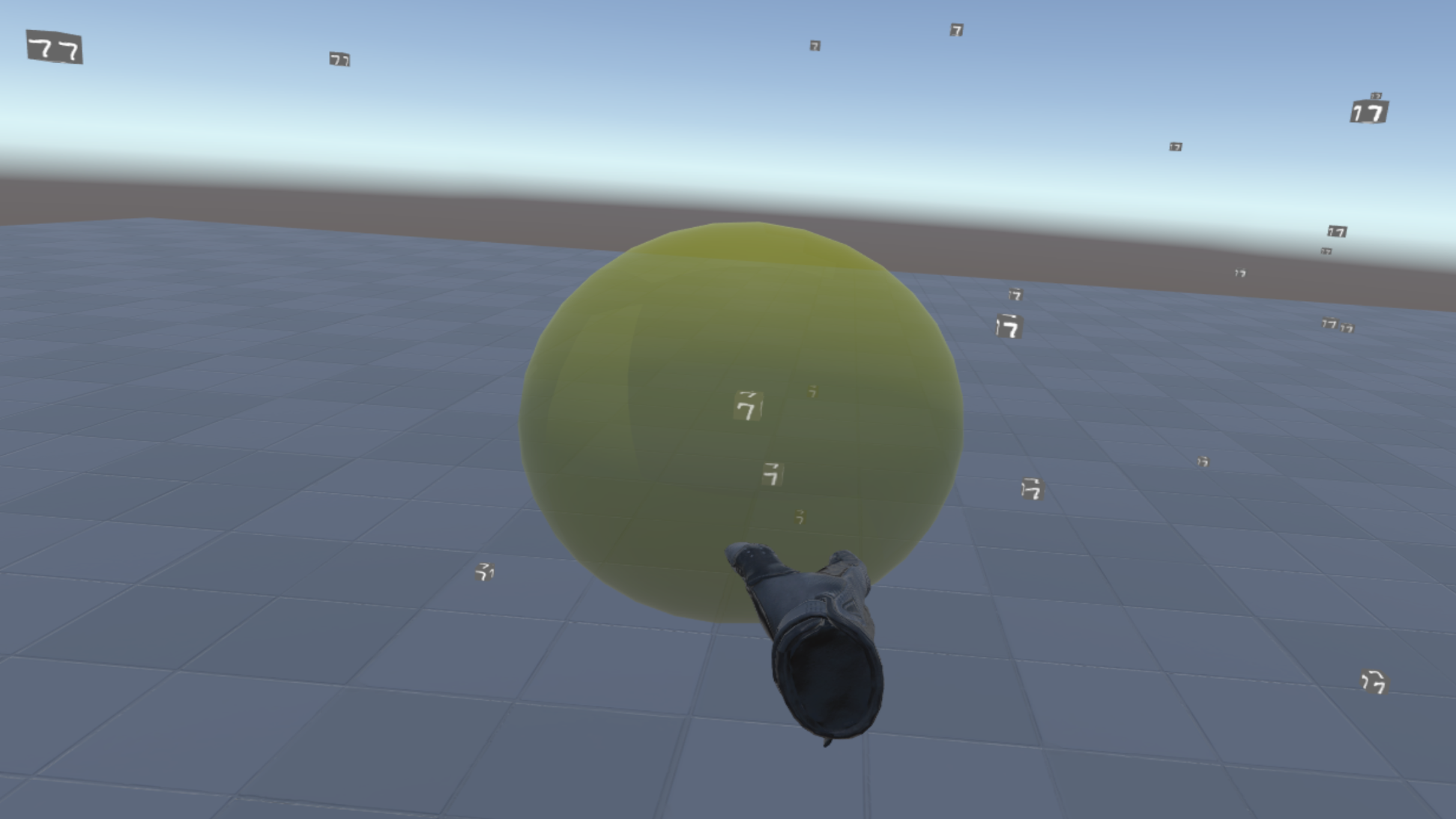}
        \caption{Selecting a label of a sphere for annotation}
        \label{fig:SphereLabel}
   \end{subfigure}
   \hspace{.0\textwidth}
   \begin{subfigure}{0.49\textwidth}
        \centering
        \includegraphics[width=\textwidth, clip]{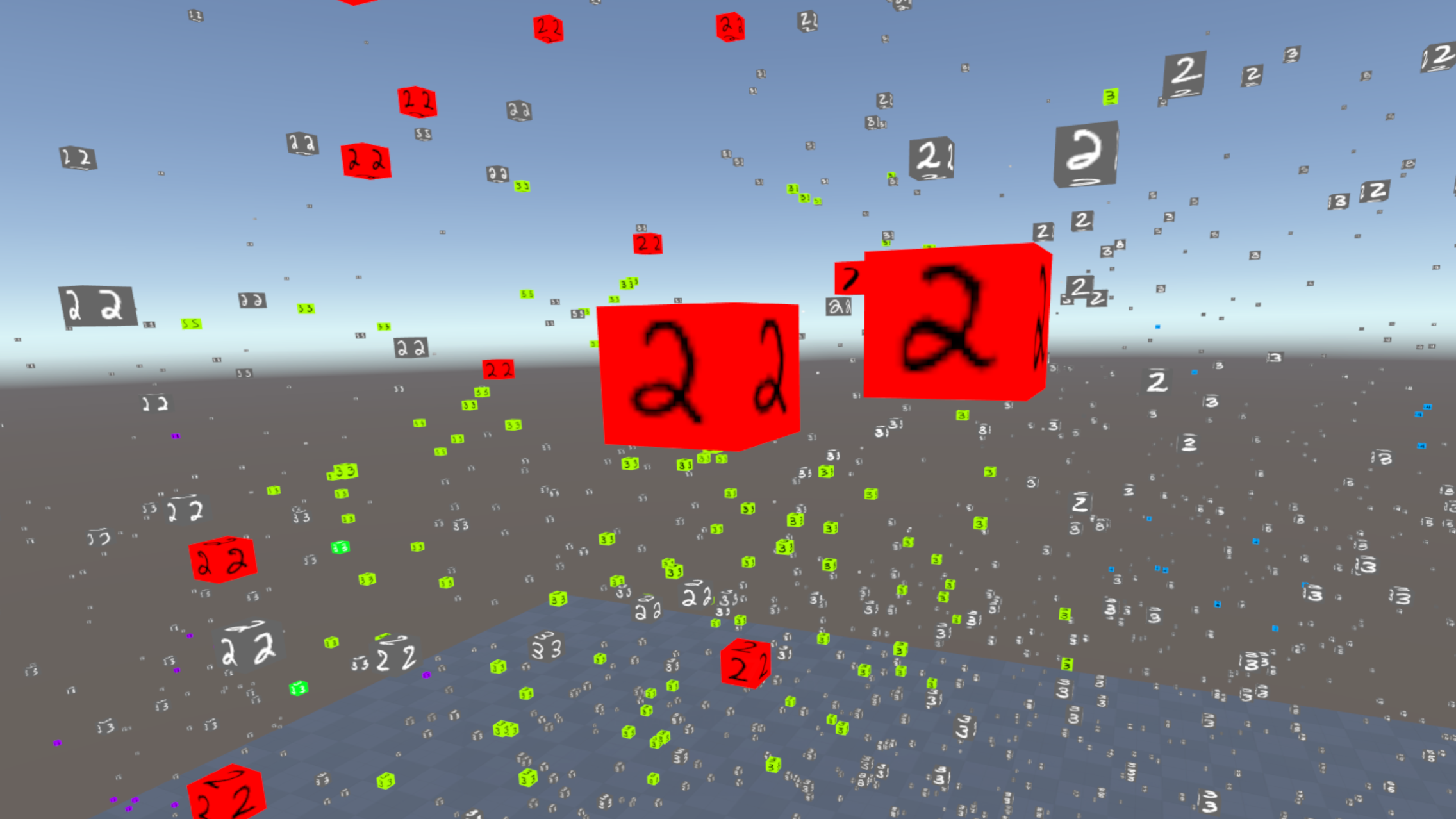}
        \caption{Dataset after labelling some data points, which appear colour-coded}
        \label{fig:DatasetFewLabelled}
   \end{subfigure}
    \caption{Steps of the annotation process from the user's point of view in the virtual space and model architecture.}
    \label{fig:scene0scene1}
\end{figure}

\section{Demonstration}
\label{sec:demonstration}

Our tool is an \gls{iui} consisting of a deep neural network linked to a \gls{vr} interface
(see \cref{sec:description} for a technical description).
An input dataset is represented as a cloud of points in virtual space;
for demonstration purposes, we use the MNIST dataset \cite{MNIST}, a standard set of images of handwritten numerals.
When entering the virtual space, the user stands outside the point cloud; this perspective offers a broad overview of the entire data set (\cref{fig:InitDataComplete}).
At this stage, the user can already notice that the points are distributed in space so that the cloud largely occupies the entire virtual space---even if not uniformly.
To get different perspectives on the data cloud the user can move in virtual space, either by physical movement or by using a teleport mechanism (\cref{fig:Teleport}).
Once inside the cloud of points, the user will see that each point is a cube, and the images (handwritten digits, in this case) are rendered as a texture on the surface of the cubes (\cref{fig:InitDataZoomed}).
Furthermore, the user might notice some degree of topological organisation in the cloud: curvy digits like 0s and 6s will be in one region of virtual space, rectilinear digits like 1s and 7s are in another region, and neighboring data points within each region tend to represent instances of the same digit.
Besides moving in virtual space, the user can also use hand gestures to assign portions of the cloud to different groups reflecting class labels (i.e. digit identity)---in other words, to annotate the data.
In the current implementation, that is done by creating and placing spheres in virtual space (\cref{fig:SpherePosition}) and assigning them a label (\cref{fig:SphereLabel}).
While non-annotated data is displayed on gray cubes, user annotated data is displayed in colored cubes indicating the assigned class (\cref{fig:DatasetFewLabelled}).
Annotating data will cause the underlying network to be updated, a process perceived by the user as motion, or gradual reshaping, of the cloud.
Motion will be perceived as if driven by three different forces: points of same category are attracted to each other, different
groups are pushed apart, and the global distribution is such that the entire space is filled.
Furthermore, the user will notice that the more data points get annotated, the more pronounced is the clustering of groups.
Importantly, data annotation reshapes the entire virtual space, and the position of each data point in virtual space is independent of whether it has been manually labelled: data points that are yet unlabelled will be spatially grouped together with labeled ones, as long as they represent similar handwritten digits.
The topological organization and motion patterns observed in virtual space are direct, tangible consequences of the way the underlying deep neural network functions.

\section{Tool Description}
\label{sec:description}  

The interactions experienced in \gls{vr}, described in \cref{sec:demonstration}, arise from the workflow presented in \cref{fig:workflow}.
Effectively, the user is annotating a dataset: the more data points are labelled, the more precisely separated the category clusters will become.
As a result, annotation efficiency is expected to gradually increase.

\begin{figure}
    \centering
    \begin{tikzpicture}
        \node[rectangle, draw, fill = {rgb:black,1;white,5}, minimum height=8em]
        (X) {\rotatebox{-90}{\textbf{Dataset}}};
        
        \node[trapezium, rotate=-90, draw, minimum width = 8em, above of = X, yshift=0.5cm, fill = white] 
        (Encoder) {\textbf{Encoder}};
        
        \node[rectangle, draw, fill = {rgb:black,1;white,5}, right of = Encoder]
        (Z) {\rotatebox{-90}{\textbf{Embedding}}};
        
        \node[trapezium, rotate=90, draw, minimum width = 8em, below of = Z, fill = white] 
        (Decoder) {\rotatebox{180}{\textbf{Decoder}}};
        
        \node[rectangle, draw, minimum height=5em, right of = Decoder, xshift =1cm, minimum height=8em]
        (Y) {\rotatebox{-90}{\textbf{Classifier}}};
    
        \begin{scope}[on background layer]
            \node[rectangle, fill = black!10, dashed, draw, text width=8em, rounded corners, inner sep = 10pt, fit=(Encoder) (Decoder), xshift = -0.cm, yshift = -0.cm, label={[shift={(0ex,0ex)}]north: Variational Autoencoder}] (DM){};
        \end{scope}
    
        \draw[->, very thick] (X.east) -- (Encoder.south);
        \draw[->, very thick] (Encoder.north) -- (Z.west);
        \draw[->, very thick] (Z.east) -- (Decoder.north);
        \draw[->, very thick] let \p1=(Y.west), \p2=(Z.south) in
        (Z.south) -- +(0,-0.6cm) -- +(2cm,-0.6cm) -- +(2cm, \y1-\y2) -- (Y.west);
    \end{tikzpicture}
    
    \caption{Basic scheme of the deep neural network architecture}
    \label{fig:modelArchitecture}

\end{figure}
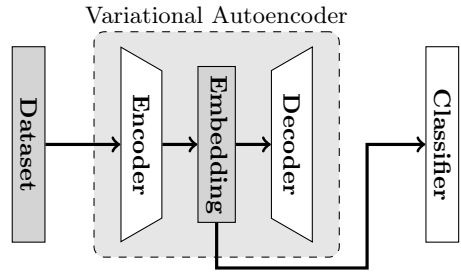

The deep neural network powering the \gls{iui} is composed of three modules: an encoder, a decoder, and a classifier (\cref{fig:modelArchitecture}).
The encoder maps input images onto a hidden layer made up of three units---in other words, it embeds images into a 3-dimensional latent representation.
The choice of the number of hidden units is not arbitrary: each unit is displayed as a dimension of virtual space.
While the encoder alone is responsible for \textit{computing} the representation of input images in virtual space, 
the other two components are essential to guide representation \textit{learning} \cite{bengio2013representation}.
The decoder maps back from 3D latent space to a reconstruction of the input image, and together with the encoder it constitutes a \gls{vae} \cite{kingma2013auto}.
The classifier, a shallow perceptron, maps from latent space onto user-provided category labels and was added to encourage cluster separation.
Following standard procedures for training neural networks, each of these tasks is expressed formally through a function measuring 
the mismatch between generated and desired outputs (objective function).
Learning takes place by iterating a two-step procedure known as gradient descent:
fist computing the direction in which network parameters should change to minimize the mismatch (i.e. the gradient),
then taking a small step in that direction.
The motions gradually reshaping the point cloud in virtual space directly reflect the iterative update of network parameters by gradient descent.
\Cref{tab:compareUserAI} establishes a direct parallel between the perspectives of the user and of the \gls{iui} system on the steps of the workflow shown in \cref{fig:workflow}.

\tikzstyle{input} = [rectangle, fill = white, draw, text width=5em, text centered, rounded corners, minimum height=3em]
\tikzstyle{block} = [rectangle, fill = white, draw, text width=12em, text centered, rounded corners, minimum height=5em]

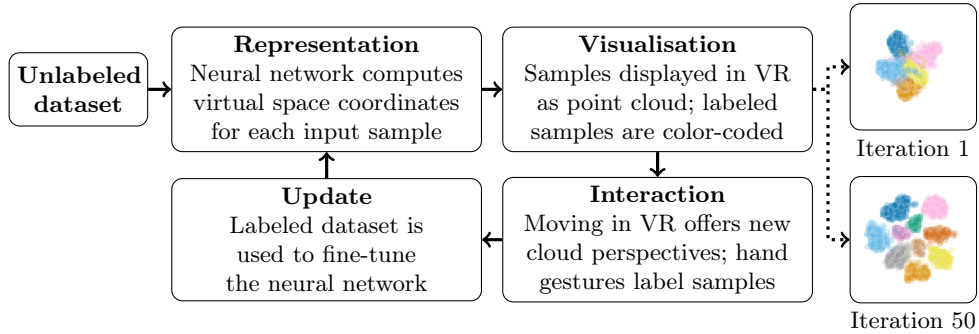
\begin{figure}[h]
    \centering

    \begin{tikzpicture}
        \node[input] (In) {\textbf{Unlabeled dataset}};
        \node[block, right of = In, xshift = 2.3cm] 
            (Model)
            {\textbf{Representation} \\
            Neural network computes virtual space coordinates for each input sample
            };
        \node[block, right of = Model, xshift = 3.4cm] 
            (VR)
            {\textbf{Visualisation} \\
            Samples displayed in \gls{vr} as point cloud; labeled samples are color-coded
            };
        \node[block, below of = VR, xshift = 0cm, yshift = -1cm] 
            (Frontend)
            {\textbf{Interaction} \\
            Moving in \gls{vr} offers new cloud perspectives;
            hand gestures label samples
            };
        \node[block, left of = Frontend, xshift = -3.4cm] 
            (Dataset2)
            {\textbf{Update} \\
            Labeled dataset is used to fine-tune the neural network
            };
        \node[rectangle, draw, rounded corners, right of = VR, xshift = 2.4cm, yshift = 0.3cm, label=below:{Iteration 1}] 
            (Data1) {\includegraphics[width=.12\textwidth]{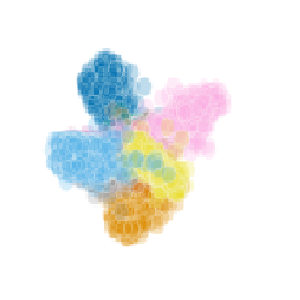}};
        \node[rectangle, draw, rounded corners, right of = VR, xshift = 2.4cm, yshift = -2cm, label=below:{Iteration 50}] 
            (Data2) {\includegraphics[width=.12\textwidth]{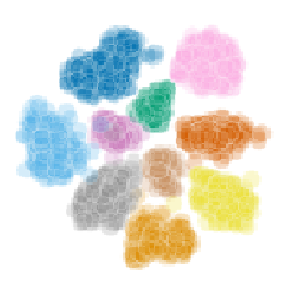}};
        \draw[->, very thick]
            (In) edge (Model)
            (Model) edge (VR)
            (VR) edge (Frontend)
            (Frontend) edge (Dataset2)
            (Dataset2) edge (Model);
        \draw[->, very thick, dotted] let \p1=(Data1.west) in (VR.east) -- +(0.2cm,0) --  +(0.2cm,\y1) -- (Data1.west);
        \draw[->, very thick, dotted] let \p1=(Data2.west) in (VR.east) -- +(0.2cm,0) -- +(0.2cm,\y1) -- (Data2.west);
    \end{tikzpicture}
   
    \caption{
        Illustration of the data labelling process. The schematic figures after 1 and 50 iterations are shown two-dimensionally and colour-coded for a better illustration of the clustering process. In our tool, the clusters are three-dimensional and only samples that are already annotated are coloured.
    }
    \label{fig:workflow}

\end{figure}

\begin{table}
    \setlength{\tabcolsep}{3pt}
    \centering
    \begin{tabularx}{\linewidth}{p{1.5cm}p{5cm}p{5cm}}
        \toprule
        State & User Perspective & Deep Learning Perspective \\
        \midrule
        
        Represen\-tation
        &
        Images from the dataset are represented as points in 3-dimensional virtual space.
        The positions of the points are not arbitrary, but show a topological organisation (e.g. curvy handwritten digits are distant from rectilinear ones, neighboring points tend to represent same digits, and the entire space is occupied).
        \vspace{0.1cm} &
        Images from the dataset are embedded in 3-dimensional latent space of a neural network. 
        The embeddings of the samples are not arbitrary, but shows a topological organisation (similar images produce similar embeddings, and global arrangement conforms to a prior distribution in latent space).
        \\
        \midrule
        Visuali\-zation
        &
        Moving in virtual space renders different perspectives on the data cloud.
        Manually annotated data points are color coded.
        \vspace{0.1cm} &
        User position in virtual space is used to compute 2-dimensional projections of 3-D embeddings without altering coordinate system.
        Manually annotated samples are associated to class labels.
        \\
        \midrule     
        Interaction
        &
        Using hand gestures in \gls{vr} (e.g. positioning a sphere around a group of data points), new data samples get annotated.
        \vspace{0.1cm} & 
        A larger fraction of data samples has associated class labels and can thus be used for supervised learning.
        \\
        \midrule
        Updating 
        &
        After the labeling is done, the data points change their position in discrete time steps. 
        Each discrete time step leads to a more accurate sorting of data points in virtual space.
        &
        After new annotations are available, the model is fine tuned on the partially annotated dataset. 
        Each iteration of the learning procedure leads to a more structured representation of the data in latent space. \vspace{1em}
        \\
        &
        The positions of the data samples are adjusted by an invisible force similar to magnetism: Samples of the same class are attracted to each other, while classes repel each other.
        This effect is also shown in \cref{fig:workflow} and makes annotation increasingly easier.        
        &
        The weights of the system for calculating the embeddings of the data samples are adjusted by a mathematical method called gradient descent: Samples of the same class produce similar embeddings, while classes are separated by a linear classifier using the annotations.
        \\
        \bottomrule
    \end{tabularx}
    \caption{Description of the four states representation, visualization, interaction and updating performed by the tool, presented from the user perspective and from the deep learning perspective.}
    \label{tab:compareUserAI}
\end{table}

\section{Discussion and Future Work}

We demonstrate an \gls{iui} tool for automating image classification in \gls{vr}.
An image dataset is represented as an actionable cloud of points that can be grouped into category classes with hand gestures.
The architecture of the underlying neural network model consists of the combination of a \gls{vae} and a shallow classifier network, and the dynamics of the network learning process are experienced as structured motion patterns in virtual space.

We chose a \gls{vr} environment as \gls{iui} framework.
In addition to cognitive and immersive aspects, the advantages of \gls{vr} over two-dimensional screens for visualization and interaction with complex data have been demonstrated in recent publications \cite{Donalek14VRvisualData, Moran2015DataVisualVR, Prange21ImageClusterVR, OlshannikovaOK015BigDataVisualisation}.
Although it has been shown that annotation of data in \gls{vr} has great potential in terms of time spent and cost, most projects prefer 2-dimensional interfaces \cite{WirthQOS19PointAtMe, kath2023PAM}.
An example of an annotation tool in \gls{vr} for labelling 3D point clouds is described in \cite{WirthQOS19PointAtMe}, and another for annotating industrial datasets using deep clustering is described in \cite{Kinga2022DataClassification}.
In order to make the workflow of the underlying model intuitive, and in line with the principles of direct manipulation \cite{Shneiderman97DirectManipulation}, we use the metaphors of space, time, and force in \gls{vr} to mediate interaction with representation and updating of the underlying neural network model.
While the use of the metaphor of interface space for representing embeddings is present in previous works \cite{Kinga2022DataClassification, Prange21ImageClusterVR}, 
the metaphors of time and force for the gradient-descent-based learning of network parameters are novel to the best of our knowledge.
As a consequence, our tool complements existing elaborations with topological organisation and dynamics that enable the annotation of multiple data samples simultaneously, thus potentially improving the efficiency of the annotation process.

We are currently teaming up with domain experts in the fields of ecology and conservation sciences interested in automating sound event detection to continue co-development of the tool presented here.
Our demo offers the opportunity for exploring variants regarding user actions and sensory representations of relevant aspects of neural network design and updating.
As next steps, we will run qualitative user studies to evaluate design alternatives such as integrating a dialog system, as well as different model architectures.
We expect that \gls{iml} tools such as the \gls{iui} illustrated here will pave the way for
empowering end users in establishing a different, more transparent relation with \gls{dl}, and accelerate the autonomous development of applications in novel areas.

\bibliographystyle{splncs04}
\bibliography{main}

\end{document}